\documentstyle[aps,preprint]{revtex}
\draft
\tighten
\begin{document}
\preprint{\vbox{\hbox{RUB-TPII-19/96}\hbox{hep-ph/9706531}}}
\title{Magnetic moments of the SU(3) decuplet baryons \\ in the chiral
quark-soliton model}
\author{Hyun-Chul Kim, Micha{\l} Prasza{\l}owicz
\footnote{Permanent address: Institute of Physics,
Jagellonian University, Cracow, Poland} and Klaus Goeke}
\address{Institute of Theoretical Physics II,\\
Ruhr-University Bochum,\\ D-44780 Bochum,\\Germany}
\date{September, 1997}
\maketitle

\begin{abstract}
Magnetic moments of baryons are studied within the chiral quark
soliton model with special emphasis on the decuplet of baryons.
The model is used to identify all symmetry breaking terms
proportional to $m_{\rm s}$. Sum rules for the magnetic moments are derived.
A ``model-independent'' analysis of the symmetry breaking terms is performed
and finally model calculations are presented, which show the importance of
the rotational $1/N_{\rm c}$ corrections for cranking of the soliton.
\end{abstract}

\section{Introduction}

The magnetic moments of the $\Delta ^{++}$ and $\Omega ^{-}$ have been
measured recently. The former one was obtained from the analysis of pion
bremstrahlung\footnote{%
Note that the result is model-dependent.} \cite{Bosshard}, $\mu _{\Delta
^{++}}=(4.52\pm 0.50)\;\mu _{\text{N}}$, while the latter one was measured by
the E756 Collaboration~\cite{E756} and found to be $\mu _{\Omega
^{-}}=(-1.94\pm 0.17\pm 0.14)\;\mu _{\text{N}}$. These experimental data have
triggered a new interest in the study of the magnetic moments of the
baryon-decuplet: For example, they have been investigated in the
relativistic quark model \cite{Schlumpf,CapstickKeister}, in chiral
perturbation theory \cite{ButlerSavageSpringer,BanerjeeMilana}, in the QCD
sum rule approach~\cite{Belyaev}, in the chiral bag model
\cite{HongBrown}, in quenched lattice gauge theory
\cite{LeinweberDraperWoloshyn}, and in the nonrelativistic quark model
with two-body exchange currents~\cite{Buchmannetal}.   
One should, however, bare in mind that
while the world average for the $\Omega^{-}$ quoted by the Particle Data
Group '96 differs only slightly from~\cite{E756}:
$\mu _{\Omega^{-}}=(-2.02\pm 0.0.05)\; \mu_{\text{N}}$,
the mean value for $\Delta^{++}$ is much less constrained: $\mu
_{\Delta^{++}}=(3.5 - 7.5)\;\mu _{\text{N}}$.

Recently, the octet magnetic moments have been studied in the chiral
quark--soliton model ($\chi $QSM; also known as the semibosonized nonlinear
Nambu--Jona-Lasinio model) ~\cite{KimPolBloGo} and it has been shown that
the $\chi $QSM reproduces the data within about $15\%$. In fact, the
accuracy that has been reached is more or less the upper limit that can be
obtained in any model with hedgehog symmetry~\cite{BaeMcGovern}. The aim of
the present work is to extend our investigation to the magnetic moments of
the baryon decuplet. Only after this work had been completed we 
were aware of Ref.\cite{WaKaya} where both octet and decuplet magnetic
moments have been calculated in a model identical to the one discussed
here. Although the results of Ref.\cite{WaKaya} are in qualitative
agreement with ours, there are some quantitative differences which we
discuss in section IV.D of the present paper. Moreover, in contrast
to the authors of Ref.\cite{WaKaya} who have contented themselves with
the model calculations, we try to present
"model-independent" analysis based on the chiral quark-soliton model,
which does not rely on dynamical calculations and does not suffer from
the ambiguities of the SU(3) quantization.

In contrast to the mass splittings,
the general analysis of the symmetry breaking
for the magnetic moments is quite involved. Famous Gell-Mann--Okubo mass
formulae are so simple, since matrix elements of the mass splitting
operator, which is assumed to transform as an octet, can be parametrized by
2 free parameters. In the case of magnetic moments there are two parameters
which parametrize magnetic moments in the chiral limit and at least 5 others
which describe chiral symmetry breaking. Under these circumstances it might
seem impossible to write general model-independent relations similar to
those for the mass splittings. Nevertheless we will adopt the strategy in
which the algebraic structure of the $\chi $QSM will serve us as a tool
to identify the relevant symmetry
breaking terms. Then the pertinent coefficients, which are of course
calculable from the solitonic profile function of the model,
will be treated as {\it free} parameters and
fitted. Despite the large number of free parameters this procedure does have
a predictive power, since the number of magnetic moments in decuplet and
octet is 18; moreover there are decays governed by the same operator which
can be described by the same set of parameters. This ``model-independent''
analysis will be at the end compared with the model predictions in line
with Ref.\cite{KimPolBloGo}, {\em .i.e.} those using the selfconsistent
profile functions.

As far as the ``model-independent'' analysis is concerned our results can be
summarized as follows: we are able to fit all but one (called $p$) free
parameters in the octet sector; this is because the system of linear
equations which we get is undetermined.
Two measured magnetic moments of $%
\Omega ^{-}$ and $\Delta ^{++}$ depend very weakly on $p$, and, as a
consequence, $p$ cannot be constrained from the data. Nevertheless, we are
able to derive sum rules, both for the octet and for the decuplet and also the
ones which involve particles from both multiplets, which to our knowledge
are new. For example we get for the magnetic moments:

\begin{equation}
\label{810sr}\Delta ^{++}-\Omega ^{-}=\frac 34\left( 2\text{p}+\text{n}%
+\Sigma ^{+}-\Sigma ^{-}-\Xi ^0-2\Xi ^{-}\right) .
\label{Eq:1}
\end{equation}
Experimentally, $\Delta^{++} - \Omega^{-} = 6.43$ with a rather large
error corresponding to the uncertainty of the world average
for $\mu_{\Delta^{++}}$ and the right hand side of Eq.(\ref{Eq:1})
$\frac 34\left( 2\text{p}+
\text{n}+\Sigma ^{+}-\Sigma ^{-}-\Xi ^0-2\Xi ^{-}\right)=7.38$.
As far as model predictions are concerned we find:
$\mu_{\Delta^{++}} =4.73$ and
$\mu_{\Omega^{-}}=-2.27$ in very good agreement with the data.

In the $\chi $QSM the baryon can be viewed as $N_{\text{c}}$ valence quarks
coupled to the polarized Dirac sea bound by a nontrivial chiral background
hedgehog field in the Hartree approximation. The proper quantum numbers of
baryons are obtained by the semi--classical quantization carried out by
integrating over zero--mode fluctuations of the pion field around the
saddle point. Hence, decuplet baryons are understood as the excited states
of a collective rotation.  In the process of quantizing the
system, the angular velocities $\Omega_a$ corresponding to the rigid rotation
of the hedgehog are promoted to the collective quantum operators, which
satisfy SU(3)$_{{\rm right}}$ (generalized spin) commutation rules. These
operators do not commute with the Wigner matrices $D_{ab}^{(\nu)}$
which naturally appear in the formalism. It has been recently shown that
inclusion of the terms proportional to the commutator $[\Omega_a,D_{bc}^{(%
\nu)}]$ which vanish classically but are nonzero in the quantum case
is of utmost importance for the phenomenological description of axial
properties of the nucleon~\cite{PraBloGo,BloPra,Waga}.
These new terms, first discussed in
Ref.\cite{WaWa}, are taken into account  in our analysis of the magnetic
moments.

One more remark should be added as far as the rotational corrections to
physical quantities are concerned\footnote{We thank the referee for
raising  this point.}. Recently \cite{Wacharge}
 it has been observed that the prescription used here (and in other works
\cite{KimPolBloGo,WaKaya,PraBloGo,kapg,BloPra}) leads in the SU(3) case
to the problem with charge quantization. Namely, because the {\em
hedgehog} is not fully SU(3) symmetric (see Eq.(\ref{U0})) some
quantities, like tensors of inertia for example, may get spurious
contributions which are totally antisymmetric. This happens only for
SU(3) and not for SU(2) because the {\em hedgehog} commutes with
$\lambda_8$ and the symmetry properties are such that terms proportional
to the structure constant $f_{8ab}$ are allowed. This is reflected in
the charge of the proton which due to these new contribution is shifted
away from 1. So far an {\em ad hoc} prescription has been adopted in
which the unwanted terms have been subtracted. It is however not clear
which terms should be subtracted in case of operators different than
electric charge; it is also not clear if all operators suffer from this
ambiguity.  Since there is no satisfactory theoretical solution to this
problem for the purpose of the present work we simply take our formulae
as they are with no subtractions. Needless to say that our
"model-independent" analysis is free from this ambiguity. This is
because it affects only the numerical values of various coefficients
predicted by the model, which in the "model-independent" analysis are taken
from the data.

The outline of the paper is as follows: In the next section, we sketch the
basic formalism for obtaining the decuplet magnetic moments in the $\chi$%
QSM. In section III, we first calculate the contribution of the leading
order and $1/N_{{\rm c}}$ rotational corrections in the chiral limit, taking
advantage of the former calculation of the octet magnetic moments. The
results are compared with the isovector relations for the magnetic moments
in the large $N_{{\rm c}}$ limit. In section IV,
we study the effects of the strange quark mass $m_{{\rm s}}$ corrections.
We also compare in section IV.D our results with those of Ref.~\cite{WaKaya}.
In the last section, we summarize the present work and draw conclusions.

\section{General Formalism}

In this section we present general formulae needed to calculate decuplet
magnetic moments (for details of the formalism see a recent review~\cite
{review}).

The $\chi $QSM is characterized by a partition function in Euclidean space
given by the functional integral over the pseudoscalar meson and quark
fields:
\begin{equation}
{\cal Z}=\int {\cal D}\Psi {\cal D}\Psi ^{\dagger }{\cal D}\pi ^a\exp {%
\left( -\int d^4x\Psi ^{\dagger }iD\Psi \right) }=\int {\cal D}\pi ^a\exp {%
(-S_{{\rm eff}}[\pi ])},\label{Eq:action}
\end{equation}
where $S_{\text{eff}}$ is the effective action
\begin{equation}
S_{\text{eff}}[\pi ]\;=\;-\mbox{Sp}\log {iD}.
\end{equation}
$iD$ represents the intrinsic Dirac differential operator
\begin{equation}
iD\;=\;\beta (-i\rlap{/}{\partial }+\hat m+MU^{\gamma _5})
\end{equation}
with the pseudoscalar chiral field
\begin{equation}
U^{\gamma _5}\;=\;\exp {(i\pi ^a\lambda ^a\gamma _5)}\;=\;\frac{1+\gamma _5}2%
U\;+\;\frac{1-\gamma _5}2U^{\dagger }.
\end{equation}
$\hat m$ is the current quark mass matrix given by
\begin{equation}
\label{Eq:mass}\hat m\;=\;\mbox{diag}(m_u,m_d,m_{{\rm s}})\;
\;=\; m_0{\bf 1} \;+\; m_8 \lambda_8.
\end{equation}
$\lambda ^a$ represent the usual Gell-Mann matrices normalized as $\mbox{tr}%
(\lambda ^a\lambda ^b)=2\delta ^{ab}$.  The $m_0$ and $m_8$ are respectively
defined by
\begin{equation}
m_0\;=\; \frac{2\overline{m}+m_{\rm s}}{3},\;\;\;\;\;m_8\;=\;
\frac{\overline{m}-m_{\rm s}}{\sqrt{3}}
\end{equation}
with isospin symmetry ($m_u+m_d = 2\overline{m}$) assumed.
$M$ stands for the dynamical quark mass arising from the
spontaneous chiral symmetry breaking, which is in general momentum-dependent~%
\cite{dp}. We regard $M$ as a constant and employ the proper-time
regularization for convenience.

The operator $iD$ is expressed in Euclidean space in terms of the Euclidean
time derivative $\partial _\tau $ and the Dirac one--particle Hamiltonian $%
h(U^{\gamma _5})$
\begin{equation}
\label{Eq:Dirac}  iD\;=\;\partial _\tau \;+\;h(U^{\gamma _5}) %
+ \beta\hat{m} - \beta\overline{m}{\bf 1}
\label{Eq:Diff}
\end{equation}
with
\begin{equation}
\label{Eq:hamil} h(U^{\gamma _5})\;=\;\frac{\vec \alpha \cdot \nabla }i%
\;+\;\beta MU^{\gamma _5}\;+\;\beta \overline{m}{\bf 1} .
\end{equation}
Here $\beta $ and $\vec \alpha $
are the well--known Dirac Hermitian matrices. $U$ is assumed to have a
structure corresponding to the embedding of the SU(2)-hedgehog into SU(3):
\begin{equation}
U\;=\;\left(
\begin{array}{cc}
U_0 & 0 \\
0 & 1
\end{array}
\right) ~~~~~{\rm with}~~~~~U_0\;=\;\exp \left( i\hat {\rm r}
\cdot \vec \tau P(r)\right) .
\label{U0}
\end{equation}
$P(r)$ is called profile function. The partition function of Eq.(\ref
{Eq:action}) can be simplified by the saddle point approximation which is
exact in the large $N_{{\rm c}}$ limit. One ends up with a stationary
profile function $P(r)$ which is evaluated by solving the Euler--Lagrange
equation corresponding to $\delta S_{{\rm eff}}/\delta P(r)=0$. This gives
the static classical field $U_0$. Note that according to Eq.(\ref{Eq:hamil})
the profile is calculated with the tail corresponding to the massive pion.
Since the masses of the up and down quarks are much smaller than
that of the strange quark, we approximate the mass term in
Eq.(\ref{Eq:Diff}) as
\begin{equation}
\hat{m} - \overline{m}{\bf 1} \; \approx\;  m_{\rm s} \left(\frac13 {\bf 1} -
\frac{1}{\sqrt{3}}\lambda^8\right).
\end{equation}

The magnetic moments of the baryon decuplet can be calculated from the
following one-current baryon matrix element:
\begin{equation}
\label{mat}\langle B | \bar{\psi}(z)\gamma_{\mu}\hat{Q} \psi(z) | B
\rangle,
\end{equation}
where $\hat{Q}$ is the charge operator of quarks in SU(3) flavor space,
defined by
\begin{equation}
\label{Eq:charge}\hat{Q}\;=\; \left(
\begin{array}{ccc}
\frac23 & 0 & 0 \\
0 & -\frac13 & 0 \\
0 & 0 & -\frac13
\end{array}
\right)\;=\; \frac12 \left(\lambda^3 + \frac{1}{\sqrt{3}}\lambda^8\right).
\end{equation}

One can relate the baryonic matrix element Eq.(\ref{mat}) to the correlation
function:
\begin{equation}
\label{corf}\langle 0|J_{B}({\bf x},T)\bar \psi \gamma _\mu \hat Q\psi
J_{B}^{\dagger }({\bf y},0)|0\rangle
\end{equation}
at large Euclidean time $T$ . The baryon current $J_B$ can be constructed
from $N_{{\rm c}}$ quark fields,
\begin{equation}
J_B=\frac 1{N_{{\rm c}}!}\varepsilon ^{i_1\ldots i_{N_{{\rm c}}}}\Gamma
_{SS_3II_3Y}^{\alpha _1\ldots \alpha _{N_{{\rm c}}}}\psi _{\alpha
_1i_1}\ldots \psi _{\alpha _{N_{{\rm c}}}i_{N_{{\rm c}}}}
\end{equation}
$\alpha _1\ldots \alpha _{N_{{\rm c}}}$ are spin--isospin indices, $%
i_1\ldots i_{N_{{\rm c}}}$ are color indices, and the matrices $\Gamma
_{SS_3II_3Y}^{\alpha _1\ldots \alpha _{N_{{\rm c}}}}$ are taken to endow the
corresponding current with the quantum numbers $SS_3II_3Y$. $%
J_B(J_B^{\dagger })$ annihilates (creates) the baryon state at given time $T$.
The rotational corrections  $1/N_{{\rm c}}$ and
linear $m_{{\rm s}}$ corrections being taken into account,
the expression for the collective magnetic moment
operator $\hat \mu$ can be written as
follows:
\begin{eqnarray}
\hat \mu  &=&\left( w_1^1+m_{{\rm s}}w_1^2\right)
\;D_{Q3}^{(8)}\;+\;w_2d_{pq3}D_{Qp}^{(8)}\cdot \hat S_q\;+
\;\frac{w_3}{\sqrt{3}}D_{Q8}^{(8)} \hat S_3  \nonumber \\
&&+m_{{\rm s}}\left[ \frac{w_4}{\sqrt{3}}d_{pq3}D_{Qp}^{(8)}D_{8q}^{(8)}+w_5%
\left( D_{Q3}^{(8)}D_{88}^{(8)}+D_{Q8}^{(8)}D_{83}^{(8)}\right)
\;+\;w_6\left( D_{Q3}^{(8)}D_{88}^{(8)}-D_{Q8}^{(8)}D_{83}^{(8)}\right)
\right] ,  \label{Eq:mu}
\end{eqnarray}
where
\begin{eqnarray}
w_1^1&=& \frac{M_N}{3} \left({\cal Q}_0
+\frac{{\cal Q}_1}{I_1}
+\frac{{\cal Q}_2}{I_2}\right), \nonumber \\
w_1^2 &= & - 2  \frac{M_N}{3} {\cal M}_0, \nonumber \\
w_2 &=& - \frac{M_N}{3} \frac{{\cal X}_2 }{I_2}, \nonumber \\
w_3 &=& - \frac{M_N}{3} \frac{{\cal X}_1 }{I_1}, \nonumber \\
w_4 &=& \frac{M_N}{3}
\left( 6 {\cal M}_2   - 2 \frac{K_2}{I_2} {\cal X}_2 \right), \nonumber\\
w_5&=&\frac{M_N}{3} \left({\cal M}_0 + {\cal M}_1 -
\frac{1}{3} \frac{K_1}{I_1} {\cal X}_1\right), \nonumber\\
w_6&=&\frac{M_N}{3} \left(
{\cal M}_0 -{\cal M}_1 +\frac{1}{3} \frac{K_1}{I_1} {\cal X}_1\right).
\label{wi}
\end{eqnarray}
The dynamical quantities $w_i$ are independent of the baryons involved. They
are expressed in terms of the inertia parameters of the soliton which have a
general structure like:
\begin{equation}
\label{spec}\sum_{m,n}\langle n|O_1|m\rangle \langle m|O_2|n\rangle {\cal R}%
(E_n,E_m,\Lambda ),
\end{equation}
where $O_i$ are spin-isospin operators changing the grand spin of states $%
|n\rangle $ by $0$ or $1$.
 The double sum runs over all the eigenstates
of the intrinsic quark Hamiltonian in the soliton field and ${\cal R}$ is
the regularization function. The numerical technique for calculating such
double sums has been developed in \cite{Blotzetal,Goetal,WaYo}. Explicit
forms of the inertia parameters are given in Appendix.  $\hat S_a$
stands for an operator of the generalized spin acting on the angular
variable $R(t)$~\cite{Blotzetal}. $D_{ab}^{(\nu)}(R)$ denote Wigner
matrix in the representation $\nu$.
 Terms ${\cal Q}_1/I_1 +{\cal Q}_2/I_2$
arise from the time-ordering of the collective operators~\cite{chretal}.

The operator (\ref{Eq:mu}) has to be sandwiched between the octet and
decuplet collective wave functions. However, strictly speaking, spin 1/2
(3/2) baryon wave functions are no longer pure octet (decuplet) states. This
is because the collective splitting Hamiltonian
\begin{eqnarray}
\hat{H^{\prime}} &=& m_{{\rm s}}\left( \alpha D_{88}^{(8)} + \beta \hat{Y} +
\frac{\gamma}{\sqrt{3}} D_{8A}^{(8)}\hat{S}_A \right)  \label{Hprime}
\end{eqnarray}
mixes the states in various SU(3) representations (here $A=1 \ldots 3$).
Constants $\alpha$, $\beta$ and $\gamma$ are given by~\cite{Blotzetal}:
\begin{equation}
\label{albega}\alpha =-\sigma+\frac{K_2}{I_2},~~~~ \beta=-\frac{K_2}{I_2}%
,~~~~ \gamma=2 \left( \frac{K_1}{I_1} - \frac{K_2}{I_2} \right).
\end{equation}
Here $K_i$ and $I_i$ are the ``moments of inertia'' and $\sigma$ is related
to the nucleon sigma term~\cite{sigma}:
$\sigma = 1/3 \; \Sigma/ \overline{m}$.

A spin $S=1/2$ state has the following form in the first order in $m_{{\rm s}%
}$:
\begin{equation}
\label{state8} \left|B,\frac 12,S_3\right\rangle
=\left|8,B,\frac 12,S_3\right\rangle+m_{\rm s}\,c_{\overline{%
10}}^B\ \left|\overline{10},B,\frac 12,S_3\right\rangle
+m_{\rm s}\,c_{27}^B\ \left|27,B,\frac 12,S_3\right\rangle
\end{equation}
and a spin $S=3/2$ state reads:
\begin{equation}
\label{state10} \left|B,\frac 32,S_3\right\rangle
=|10,B,\frac 32,S_3\rangle
+m_{{\rm s}}\,a_{27}^B\ \left|27,B,\frac 32,S_3\right\rangle
+m_{{\rm s}}\,a_{35}^B\ \left|35,B,\frac 32,S_3\right\rangle,
\end{equation}
where
\begin{equation}
\label{aaB}c_{\overline{10}}^B=c_{\overline{10}}\left[
\begin{array}{c}
\sqrt{5} \\ 0 \\
\sqrt{5} \\ 0
\end{array}
\right] ,\;c_{27}^B=c_{27}\left[
\begin{array}{c}
\sqrt{6} \\ 3 \\
2 \\
\sqrt{6}
\end{array}
\right] ,\;a_{27}^B=a_{27}\left[
\begin{array}{c}
\sqrt{15/2} \\ 2 \\
\sqrt{3/2} \\ 0
\end{array}
\right] ,\;a_{35}^B=a_{35}\left[
\begin{array}{c}
5/
\sqrt{14} \\ 2
\sqrt{5/7} \\ 3
\sqrt{5/14} \\ 2\sqrt{5/7}
\end{array}
\right]
\end{equation}
in the basis $[N,\Lambda ,\Sigma ,\Xi ]$ and $[\Delta ,\Sigma ^{*},\Xi
^{*},\Omega ]$ respectively, and
\begin{eqnarray}
c_{\overline{10}}=\frac{I_2}{15}\left( \sigma -\frac{I_1}{K_1}\right) ,
&~&c_{27}=\frac{I_2}{25}\left( \sigma +\frac{I_1}{3K_1}-\frac{4I_2}{3K_2}%
\right) ,  \nonumber \\
~~ &&  \nonumber \\
a_{27}=\frac{I_2}8\left( \sigma -\frac{5K_1}{3I_1}+\frac{2K_2}{3I_2}\right)
, &&a_{35}=\frac{I_2}{24}\left( \sigma +\frac{K_1}{I_1}-\frac{2K_2}{I_2}%
\right) .  \label{ca}
\end{eqnarray}

In actual calculations we shall need the explicit form of the SU(3) wave
functions corresponding to the states of Eqs.(\ref{state8},\ref{state10}).
Let us remind that the wave function of a state of flavor $B=(Y,T,T_3)$ and
spin $S=(Y^{\prime }=-1,S,S_3)$ in the representation $\nu$ is given in
terms of a tensor with two indices:
$\psi _{(\nu\,B),(\overline{\nu}\,\overline{S})}$,
one running over the states $B$ in the representation
$\nu$ and the other one over the states
$\overline{S}$ in the representation $\overline{\nu}$.
  Here $\overline{\nu}$ denotes the complex
conjugate of the representation $\nu$,
and the conjugate of the state $S$ is
given by: $\overline{S}=(1,S,-S_3)$\footnote{Here we use $S$ to
denote not only the spin but also the corresponding SU(3) state
with hypercharge $-1$. It is always clear from the context how $S$ should be
understood.}. Explicitly~\cite{Blotzetal}:
\begin{equation}
\label{psi}\psi _{(\nu\,{\cal B}),
(\overline{\nu}\,\overline{\cal S})}=\sqrt{{\rm dim}%
(\nu)}\,(-)^{Q_S}\left[ D_{BS}^{(\nu)}\right] ^{{\bf *}},
\end{equation}
where $Q_S$ is a charge corresponding to the SU(3) state $S$. The explicit
dependence on the SU(3) rotation matrix has been suppressed in Eq.(\ref{psi}%
).

Now, by sandwiching (\ref{Eq:mu}) between the states
(\ref{state8},\ref{state10})
we get the following expression for the magnetic moments with ${\cal O}(m_{%
{\rm s}})$ accuracy:
\begin{equation}
\label{mB}\mu (B)=\mu _0(B)+m_{{\rm s}}\,\mu _1^{({\rm op)}}(B)+m_{{\rm s}%
}\,\mu _1^{({\rm wf)}}(B),
\end{equation}
where by $\mu _0$ we have denoted the chiral limit part of the magnetic
moment, $\mu _1^{({\rm op})}$ comes from the symmetry breaking in the
magnetic moment operator ({\em i.e.} from terms proportional to $w_4,w_5$
and $w_6$) and $\mu _1^{({\rm wf})}$ arises from the interference between
the ${\cal O}(m_{{\rm s}})$ and ${\cal O}(1)$ parts of the wave functions
(\ref{state8},\ref{state10}) with $\mu _0$. It is quite straightforward to
evaluate the SU(3) matrix elements required in order to calculate 
(\ref{mB}). One has, however, to remember that the wave 
functions (\ref{psi}) by construction
transform under the transformations generated by the operators $\hat S$ as
tensors in the representation $\overline{\nu}$ rather than $\nu$.
Then using the known formulae for integrating the products of the Wigner $%
D^{(\nu)}$ functions over the SU(3) group and SU(3) coupling
coefficients~\cite{KW}, one gets for the octet $[N,\Lambda ,\Sigma ,\Xi ]$:
\begin{eqnarray}
\mu _0(B_8) &=&-\frac 1{30}\left[
\begin{array}{c}
14T_3+1 \\
-3 \\
5T_3+3 \\
-4T_3-4
\end{array}
\right] \left( w_1^1-\frac 12w_2\right) \,S_3+\frac 1{60}\left[
\begin{array}{c}
2T_3+3 \\
1 \\
5T_3-1 \\
8T_3-2
\end{array}
\right] \,w_3\,S_3,  \label{m0B8} \\
&&  \nonumber \\
\mu _1^{({\rm op)}}(B_8) &=&-\frac 1{270}\left[
\begin{array}{c}
22T_3-3 \\
9 \\
9T_3-5 \\
-4T_3+6
\end{array}
\right] \,w_4\,S_3+\frac 1{45}\left[
\begin{array}{c}
-10T_3 \\
0 \\
-3T_3+2 \\
4T_3-3
\end{array}
\right] \,w_5\,S_3-  \nonumber \\
&&\frac 1{15}\left[
\begin{array}{c}
2T_3 \\
0 \\
-T_3 \\
2T_3
\end{array}
\right] \,w_6\,S_3-\frac 1{30}\left[
\begin{array}{c}
14T_3+1 \\
-3 \\
5T_3+3 \\
-4T_3-4
\end{array}
\right] \,w_1^2\,S_3,  \label{m1opB8} \\
&&  \nonumber \\
\mu _1^{({\rm wf)}}(B_8) &=&-\frac 13\,c_{\overline{10}}\left[
\begin{array}{c}
2T_3-1 \\
0 \\
T_3-1 \\
0
\end{array}
\right] \,\left( w_1+w_2+\frac 12w_3\right) \,S_3+  \nonumber \\
&&-\frac 1{45}\,c_{27}\left[
\begin{array}{c}
4T_3+6 \\
9 \\
4 \\
-4T_3+6
\end{array}
\right] \,\left( w_1+2w_2-\frac 32w_3\right) \,S_3,  \label{m1wfB8}
\end{eqnarray}
and for the decuplet $[\Delta ,\Sigma ^{*},\Xi^{*},\Omega ]$:
\begin{eqnarray}
\mu _0(B_{10}) &=&-\frac 1{12}\,\left( w_1^1-\frac 12w_2-\frac 12w_3\right)
Q\,S_3,  \label{m0B10} \\
&&  \nonumber \\
\mu _1^{({\rm op)}}(B_{10}) &=&-\frac 1{252}\left( \frac 13\left[
\begin{array}{c}
11Q-13 \\
15Q+2 \\
19Q+17 \\
-9Q
\end{array}
\right] \,w_4+2\left[
\begin{array}{c}
5Q-4 \\
3Q-1 \\
Q+2 \\
-6Q
\end{array}
\right] \,w_5\right) \,S_3-\frac 1{12}w_1^2Q\,S_3,  \label{m1opB10} \\
&&  \nonumber \\
\mu _1^{({\rm wf)}}(B_{10}) &=&-\frac 1{36}\,a_{27}\left[
\begin{array}{c}
5Q-10 \\
6Q-4 \\
7Q+2 \\
0
\end{array}
\right] \,\left( w_1+\frac 12w_2+\frac 32w_3\right) \,S_3+  \nonumber \\
&&-\frac 1{84}\,a_{35}\left[
\begin{array}{c}
1 \\
2 \\
3 \\
4
\end{array}
\right] \,(Q+2)\,\left( w_1+\frac 52w_2-\frac 52w_3\right) \,S_3.
\label{m1wfB10}
\end{eqnarray}
Here $Q$ is the charge and $T_3$ the third component of the isospin of the
baryon $B$ while $S_3$ is its spin projection on the third axis.

A remark concerning Eqs.(\ref{m0B10}, \ref{m1opB10}) and (\ref{m1wfB10}) is
in order. In the chiral limit one gets a simple formula in which magnetic
moments in decuplet are proportional to the corresponding
electric charge. This simple
proportionality is broken by the ${\cal O}(m_{{\rm s}})$ corrections, both
for the ``operator'' and the wave function contributions. Note also that due
to the symmetry properties of the Clebsch-Gordan coefficients there is no
contribution proportional to $w_6$ for the decuplet. On the contrary, $w_6$
does contribute in the octet case, because of the interference between $8_{%
{\rm a}}$ and $8_{{\rm s}}$ SU(3) representations.

\section{Magnetic moments in the chiral limit}

In the case of the chiral limit ($m_{{\rm s}}=0$), {\em i.e.} with $U$-spin
symmetry unbroken, we can relate the decuplet magnetic moments to those of
the octet baryons. We introduce two parameters consisting of
$w_1$, $w_2$ and $w_3$:
\begin{equation}
\label{vw}v=\frac 1{60}\left( w_1-\frac 12w_2\right) ,~~w=\frac 1{120}\;w_3,
\end{equation}
with\footnote{When we go off the chiral limit, for the purpose of the
``model-independent'' analysis we include the
correction proportional to $w_1^2$ in the definition of $w_1$.}
$w_1=w_1^1 +m_{\rm s}w_1^2$.
Using these two parameters, we can express the octet and decuplet magnetic
moments as follows:
$$
\mu _p=\mu _{\Sigma ^{+}}=-8v+4w,
$$
$$
\mu _n=\mu _{\Xi ^0}=6v+2w,
$$
$$
\mu _\Lambda =-\mu _{\Sigma ^0}=3v+w,
$$
\begin{equation} \label{mom8}
\mu _{\Sigma ^{-}}=\mu _{\Xi ^{-}}=2v-6w
\end{equation}
and for the decuplet:
\begin{equation}
\label{Eq:decuplet}
\mu _{B}=\;-\frac{15}2\left( v-w\right) \, Q_B.
\end{equation}
Here $Q_{B}$ denotes the charge of the baryon $B$.

In addition to the $U$-spin symmetry, one can also see that the generalized
Coleman and Glashow sum rules are satisfied in the chiral limit~\cite
{ColemanGlashow,HongBrown} 
\begin{eqnarray}
\mu _{\Sigma ^{*0}} &=&\frac 12\left( \mu _{\Sigma ^{*+}}+\mu _{\Sigma
^{*-}}\right) ,  \nonumber \\
\mu _{\Delta ^{-}}+\mu _{\Delta ^{++}} &=&\mu _{\Delta ^0}+\mu _{\Delta
^{+}},  \nonumber \\
\sum_{B \in {\rm decuplet}}\mu_B &=&0.  \label{Eq:ColGl}
\end{eqnarray}

In principle one can fit $v$ and $w$ to the experimental data within the
octet and then make predictions for the decuplet. However, since the $U$%
-spin symmetry is rather strongly broken in the real world,
one cannot expect this procedure
to be quantitatively accurate. Indeed:%
\begin{equation}
\begin{array}{ccccccc}
20v & = & 2\mu _{\text{n}}-\mu _{\text{p}} & = & 3\mu _{\Xi ^0}+\mu _{\Xi
^{-}} & = & -2\mu _{\Sigma ^{-}}-3\mu _{\Sigma ^{+}} \\
&  & (-6.61) &  & (-4.40) &  & (-5.06)
\end{array}
\label{v}
\end{equation}
and for $w$:%
\begin{equation}
\begin{array}{ccccccc}
20w & = & 4\mu _{\text{n}}+3\mu _{\text{p}} & = & \mu _{\Xi ^0}-3\mu _{\Xi
^{-}} & = & -4\mu _{\Sigma ^{-}}-\mu _{\Sigma ^{+}}, \\
&  & (0.73) &  & (0.70) &  & (2.18)
\end{array}
\label{w}
\end{equation}
where numbers in parenthesis correspond to the experimental values. One can
observe very strong disagreement between the theoretical
formulae in the chiral limit and the experiment,
especially for $w$. This means, of course, that the
sum rules of Eqs.(\ref{v},\ref{w}) will be strongly violated by $m_{\text{s}%
} $ corrections. As will be shown in the next section only the {\it mean}
values of the three terms contributing to $v$ or $w$
in Eqs.(\ref{v},\ref{w}) are free of the $m_{%
\text{s}}$ corrections. With this in mind we get:%
\begin{equation}
\begin{array}{cc}
v=-0.268\,, & w=0.060\;.
\end{array}
\label{vwnum}
\end{equation}
Had we used in Eqs.(\ref{v},\ref{w})
the nucleon or the $\Xi $ data alone to fit $w$
the result would be a factor of 2 smaller.

Using Eq.(\ref{vwnum}) we obtain the octet magnetic moments:
$$
\mu _p\;(2.79)=\mu _{\Sigma ^{+}}\;(2.46)=2.38,
$$
$$
\mu _n\;(-1.91)=\mu _{\Xi ^0}\;(-1.25)=-1.49,
$$
$$
\mu _\Lambda \;(-0.61)=-\mu _{\Sigma ^0}\;(\mbox{no data}) =-0.74,
$$
$$
\mu _{\Sigma ^{-}}\;(-1.16)=\mu _{\Xi ^0}\;(-1.25)=-0.90
$$
and for the decuplet:
$$
\mu _{\text{B}}=2.46\;Q_{\text{B}}
$$
(in units of nuclear magneton $\mu _{\text{N}})$. In particular:%
$$
\begin{array}{ccc}
\mu _{\Delta ^{++}}\;(4.52)=4.92\; & \text{and} & \mu _{\Omega
^{-}}\;(-1.94)=-2.46,
\end{array}
$$
where numbers in parenthesis correspond to the experimental data.

In summary let us note that with theoretical formulae
derived from the $\chi $QSM in
the chiral limit with ${\cal O}(1/N_{\text{c}})$ corrections 
we are able to describe data with accuracy of the order $25-30\ \%$
as far as octet baryons are concerned.  The same concerns decuplet
magnetic moments where, however, the
experimental situation is less clear. It is therefore evident that large
corrections due to the strange quark mass are expected.

One more remark is here in order: although in the chiral limit magnetic moments
depend on 3 model parameters, namely $w_1$, $w_2$ and $w_3$, only the
combination $w_1-1/2\;w_2\sim v$ enters. As we shall see in the next section
it will be possible to extract $w_1$ and $w_2$ separately only through the
wave function corrections (see Eq.(\ref{mB})).

Let us now turn to the results calculated within the $\chi $QSM
by a selfconsistent solution of the equation of motion for the solitonic
profile function.
In Table I the numerical results are presented,
as we vary the constituent quark mass $M=370,\;400,\;420,\;
\mbox{and}\;450$ MeV. It is found that as the
constituent quark mass increases the magnetic moments in general decrease.
Once more the importance of the ${\cal O}(1/N_{\text{c}})$ corrections can
be observed.

It is interesting to extract model predictions for $v$ and $w$ (for $M=420$
MeV):%
\begin{equation}
\begin{array}{cc}
v^{\text{NJL}}=-0.269\,, & w^{\text{NJL}}=0.029 .
\end{array}
\label{vwNJL}
\end{equation}
One can see from Eq.(\ref{vwNJL}) that the model prediction for $w$ is 2
times smaller than the estimate of our ``model-independent'' analysis. As
mentioned after Eq.(\ref{vwnum}) this value (\ref{vwNJL}) of $w$ is compatible
with the nucleon and $\Xi $ magnetic moments but contradicts
the data for the $\Sigma $.

In the large $N_{{\rm c}}$ limit, isovector magnetic moments satisfy the
following relations~\cite{JenMan}
\begin{eqnarray}
\frac{\mu_{\Delta^{++}} - \mu_{\Delta^{-}}} {\mu_{p} - \mu_{n}} & = & \frac95
+ {\cal O} (\frac{1}{N_{{\rm c}}^{2}}),  \nonumber \\
\frac{\mu_{\Delta^{+}} - \mu_{\Delta^{0}}} {\mu_{p} - \mu_{n}} & = & \frac35
+ {\cal O} (\frac{1}{N_{{\rm c}}^{2}}),  \nonumber \\
\frac{\mu_{\Sigma^{*+}} - \mu_{\Sigma^{*-}}} {\mu_{\Sigma^{+}} -
\mu_{\Sigma^{-}}} & = & \frac32 + {\cal O} (\frac{1}{N_{{\rm c}}^{2}}),
\nonumber \\
\frac{\mu_{\Xi^{*0}} - \mu_{\Xi^{*-}}} {\mu_{\Xi^{0}} - \mu_{\Xi^{-}}} & = &
-3 + {\cal O} (\frac{1}{N_{{\rm c}}^{2}}).  \label{Eq:Ncbehave}
\end{eqnarray}
It is interesting to see if the present results for $N_c=3$
satisfy these relations.  In
fact, our model is exact in the large $N_{{\rm c}}$ limit as we mentioned
before.  Our results in the chiral limit for $N_c=3$ agree with
Eq.(\ref{Eq:Ncbehave}) within $1\sim 4\%$ except for the relation
$(\mu _{\Xi ^{*0}}-\mu _{\Xi ^{*-}})/(\mu _{\Xi ^0}-\mu _{\Xi ^{-}})$, from
which our result deviates by around $12\%$. This indicates that the
extrapolation from the large $N_{{\rm c}}$ to $N_{{\rm c}}=3$
for the above relation is well justified.
\section{Magnetic moments for the finite strange quark mass}

\subsection{Mass splittings revisited}

In order to proceed with the ``model-independent'' analysis we have to
estimate mixing parameters $c_i$ and $a_i$ defined in
Eqs.(\ref{aaB},\ref{ca}). For the
purpose of this analysis we take the following values for the physical
parameters:
\begin{equation}
\label{par}\Sigma=48\;{\rm MeV}, ~~~ m_{{\rm s}}=180\,{\rm MeV}, ~~~
\overline{m}=6\,{\rm MeV}.
\end{equation}
With this set of parameters:
\begin{equation}
\label{par1}\sigma = 2.67.
\end{equation}

In order to estimate the ratios $K_i/I_i$ entering Eqs.(\ref{ca}) we observe
that:
\begin{equation}
M_{\Sigma}-M_{\Lambda} = m_{{\rm s}} \left( \frac{1}{5} \sigma -\frac{3}{5}%
\frac{K_1}{I_1} +\frac{2}{5}\frac{K_2}{I_2} \right) = 77\, {\rm MeV}
\end{equation}
and
\begin{equation}
M_{\Xi}-M_{{\rm N}} = m_{{\rm s}} \left( \frac{1}{2} \sigma +\frac{1}{2}%
\frac{K_1}{I_1} +\frac{K_2}{I_2} \right) = 379 \, {\rm MeV}.
\end{equation}
From the above equations one obtains:
\begin{equation}
\frac{K_1}{I_1}=0.51, ~~~~ \frac{K_2}{I_2}=0.52.
\end{equation}
For the purpose of the ``model-independent'' analysis it is convenient to
make use of the approximate equality of the ratios:
$$
\frac{K_1}{I_1} \approx \frac{K_2}{I_2}.
$$
Then:
\begin{equation}
\label{approx}m_{{\rm s}} c_{\overline{10}}=c,~~ m_{{\rm s}} c_{27}=\frac{3}{%
5}\,c,~~ m_{{\rm s}} a_{27}=\frac{15}{8}\, c,~~ m_{{\rm s}} a_{35}=\frac{15}{%
24}\, c,
\end{equation}
where
\begin{equation}
c=0.14 \; m_{{\rm s}} I_2.
\end{equation}

Unfortunately $I_2$ cannot be constrained from the mass splittings without
making further assumptions about masses of the baryons belonging to higher
SU(3) representations entering Eqs.(\ref{state8},\ref{state10}).
In fact $I_2$ is responsible {\it e.g.} for the octet
antidecuplet splitting. There are theoretical predictions~\cite{Anti10}
that the lightest member of antidecuplet should have a mass of the order
of 1530 MeV.  That would mean that $I_2 \sim 0.8$ fm;
on the other hand the model predictions are somewhat smaller:
$I_2\sim 0.5-0.7$ fm depending on the constituent mass $M$.
In Ref.~\cite{Anti10}, which discusses a possible physical interpretation
of the antidecuplet it is suggested that $I_2\sim 0.4\ {\rm fm}$.
 For the purpose of further analysis we
leave $c$ (or equivalently $I_2$) as a free parameter.

\subsection{Magnetic moments of the octet}

Let us define a new set of parameters:
$$
x = \frac{1}{540} m_{{\rm s}} w_4,~~ y =\frac{1}{90} m_{{\rm s}}
w_5,~~ z = \frac{1}{30} m_{{\rm s}} w_6,~~
$$
\begin{equation}
p =\frac{1}{6} c \,\left( w_1 + w_2 +\frac{1}{2} w_3 \right),~~ q=-\frac{1}{%
150} c \, \left( w_1 + 2 w_2 -\frac{3}{2} w_3 \right).
\label{Eq:pq}
\end{equation}
In this notation Eqs.(\ref{m0B8},\ref{m1opB8},\ref{m1wfB8})
read:
\begin{equation}
\label{octetm}
\left[
\begin{array}{rrrrrrr}
-8 & 4 & -8 & -5 & -1 & \;0 & \;8 \\
6 & 2 & 14 & 5 & 1 & 2 & 4 \\
3 & 1 & -9 & 0 & 0 & 0 & 9 \\
-8 & 4 & -4 & -1 & 1 & 0 & 4 \\
2 & -6 & 14 & 5 & -1 & 2 & 4 \\
6 & 2 & -4 & -1 & -1 & 0 & 4 \\
2 & -6 & -8 & -5 & 1 & 0 & 8
\end{array}
\right] \left[
\begin{array}{c}
v \\
w \\
x \\
y \\
z \\
p \\
q
\end{array}
\right] =\left[
\begin{array}{r}
2.79 \\
-1.91 \\
-0.61 \\
2.46 \\
-1.16 \\
-1.25 \\
-0.65
\end{array}
\right] ;\;\;\left[
\begin{array}{l}
\text{p} \\ \text{n} \\ \Lambda ^0 \\
\Sigma ^{+} \\
\Sigma ^{-} \\
\Xi ^0 \\
\Xi ^{-}
\end{array}
\right] .
\end{equation}
Note that: $x,y,z,p,q\sim m_{\text{s}}$. This set of equations is not
linearly independent; indeed there is one null vector corresponding to the
following sum rule derived already in Ref.\cite{HongBrown}.
$$
12\,{\rm p} + 7 \,{\rm n} -7\, \Sigma^- + 22\, \Sigma^+ 12\,+ \Lambda
+ 3\, \Xi^+ + 23\, \Xi^0=0,
$$
where symbols denoting particles stand for their corresponding
magnetic moments.

In order to solve (\ref{octetm}) we first observe that it is possible to
find linear combinations of equations (\ref{octetm}) which give
\begin{equation}
\label{vwz}
\begin{array}{ccrcc}
v & = & -0.268 & = & \left( 2
\text{n}-\text{p}+3\Xi ^0+\Xi ^{-}-2\Sigma ^{-}-3\Sigma ^{+}\right) /60, \\ w
& = & 0.060 & = & \left( 3
\text{p}+4\text{n}+\Xi ^0-3\Xi ^{-}-4\Sigma ^{-}-\Sigma ^{+}\right) /60, \\ z
& = & -0.080 & = & (\text{n}-\Sigma ^{-}+\Sigma ^{+}-\Xi ^0-\text{p}+\Xi
^{-})\,/6,
\end{array}
\end{equation}
independently of $x,~y,~p$ and $q$. As we see from Eq.(\ref{vwz}) $w$ is
free of $m_{{\rm s}}$ corrections only if we take the average of the three
expressions given by Eq.(\ref{w}). This value of $w$ is larger as the one
predicted by the $\chi$QSM and this will have consequence for the
phenomenological predictions.

Since the set of equations (\ref{octetm}) is linearly dependent we can solve
it only as a function of one free parameter which we choose to be $p$. Before
we quote results for $w_i$ let us note that the solution for $q$ reads:
\begin{equation}\label{q}
 q=0.0118-0.1111\;p.
\end{equation}
The results for $w_i$ read:
\begin{equation}
\label{wi2}
\begin{array}{ccl}
w_1 & = & -13.76
\frac{0.229 - \, p}{0.265 - \, p}, \\
w_2 & = & 4.63
\frac{0.475 - \, p}{0.265 - \, p}, \\
w_3 & = & 7.22, \\
w_4 & = & -14.20 - 60.0 \, p, \\
w_5 & = & -0.40, \\
w_6 & = & -2.40.
\end{array}
\end{equation}
This set of $w_i$ corresponds to
\begin{equation}
c = 0.29 - 1.09 \, p ~~~{\rm and}~~~ I_2 [{\rm fm}]= 2.19 - 8.28 \, p.
\end{equation}
>From our discussion at the end of section IV.A we see that for
$I_2\sim 0.4-0.8 \ {\rm fm}$
\begin{equation}\label{plimits}
p = 0.17 - 0.22,%
\end{equation}
however, strictly speaking, we have no handle to fix $p$ from the octet
magnetic moments alone. Therefore, as already said,
we leave $p$ as a free parameter for the time being.

Before proceeding to the discussion of the decuplet
let us observe that with the set of parameters of Eqs.(\ref{q},\ref{wi2}):
\begin{equation}
\Sigma_8=\sum_{B\in {\rm octet}}\mu _{B}
= 0.53 ~~~({\rm exp}.~~0.35).
\end{equation}
It is important to observe that this sum is entirely given in terms of the
wave function corrections, it is namely equal:
\begin{equation}
\sum_{B\in {\rm octet}}\mu _{B} = 5p+45q
\end{equation}
which  in view of Eq.(\ref{q}) is independent of $p$.
In contrast to the $\chi$QSM or the Skyrme Model our ``model-independent''
analysis gives positive number for $\Sigma_8$. This is precisely due to the
larger value for $w\sim w_3$ which is required by our fitting procedure.

\subsection{``Model-independent'' analysis of decuplet magnetic moments}

Before applying the results of the previous section to the decuplet case let
us first discuss sum rules which result from Eqs.(\ref{m0B10},\ref{m1opB10},%
\ref{m1wfB10}) which can be conveniently rewritten in a following form:
\begin{equation}
\left[
\begin{array}{r}
-2 \\
-1 \\
0 \\
1 \\
-1 \\
0 \\
1 \\
0 \\
1 \\
1
\end{array}
\right] \,t+\left[
\begin{array}{r}
-9 \\
2 \\
13 \\
24 \\
-17 \\
-2 \\
13 \\
-17 \\
2 \\
-9
\end{array}
\right] \xi +\left[
\begin{array}{r}
-6 \\
-1 \\
4 \\
9 \\
-2 \\
1 \\
4 \\
-2 \\
-1 \\
-6
\end{array}
\right] \eta +\left[
\begin{array}{r}
0 \\
5 \\
10 \\
15 \\
-2 \\
4 \\
10 \\
-2 \\
5 \\
0
\end{array}
\right] \,r+\left[
\begin{array}{r}
-4 \\
-3 \\
-2 \\
-1 \\
-6 \\
-4 \\
-2 \\
-6 \\
-3 \\
-4
\end{array}
\right] \,s=\left[
\begin{array}{l}
\Delta ^{++} \\
\Delta ^{+} \\
\Delta ^0 \\
\Delta ^{-} \\
\Sigma ^{*+} \\
\Sigma ^{*0} \\
\Sigma ^{*-} \\
\Xi ^{*0} \\
\Xi ^{*-} \\
\Omega ^{-}
\end{array}
\right], \label{decup}
\end{equation}
where
$$
t=\frac 18\left( w_1-\frac 12w_2-\frac 12w_3\right) =\frac{15}2\left(
v-w\right) ,
$$
$$
r=\frac 5{64}\,c\,\left( w_1+\frac 12w_2+\frac 32w_3\right) ,
$$
$$
s=\frac 5{448}\,c\,\left( w_1+\frac 52w_2-\frac 52w_3\right) ,
$$
\begin{equation}
\xi =\frac{15}{14}\,x=\frac 1{504}m_{\text{s}}w_4,\ \qquad \eta =\frac{15}{14%
}\,y=\frac 1{84}m_{\text{s}}w_5. \label{trsxieta}
\end{equation}
Equation (\ref{decup}) without wave function corrections, {\it i.e.} with $%
r=s=0$ is identical to the one derived in Ref.\cite{HongBrown}.
The authors of Ref.\cite{HongBrown}  diagonalize the splitting
Hamiltonian (\ref{Hprime})
up to all orders in $m_{\text{s}}$ via the procedure introduced by Yabu
and Ando~\cite{YabuAndo}.
 Our procedure is perturbative in $m_{\text{s}}$ and as a
result we are able to study analytically rather than only numerically the
influence of the wave function corrections on the magnetic moments.
Moreover, similarly to the octet case, we are able to derive sum
rules for the decuplet magnetic moments. Note that in the chiral limit ($%
x=y=r=s=0$) the sum of magnetic moments is proportional to the sum of
charges\footnote{One should keep in mind that $t$ is also $m_{\rm s}$
dependent through $w_1^2\; m_{\rm s}$; see Eq.(\ref{Eq:mu}).}, which is
a simple consequence of the fact that the first column of
Eq.(\ref{decup}) is simply $-Q$ (charge). For the finite $m_{\text{s}}$
only some of these sum rules survive:%
$$ -4\Delta ^{++}+6\Delta^{+}+3\Sigma ^{*+}-6\Sigma ^{*0}+\Omega ^{-}=0, $$
$$ -2\Delta^{++}+3\Delta ^{+}+2\Sigma ^{*+}-4\Sigma ^{*0}+\Xi ^{*-}=0, $$
$$-\Delta ^{++}+2\Delta ^{+}-2\Sigma ^{*0}+\Xi ^{*0}=0, $$
$$\Sigma ^{*+}-2\Sigma ^{*0}+\Sigma ^{*-}=0,$$
$$2\Delta ^{++}-3\Delta ^{+}+\Delta^{-}=0,$$
\begin{equation}
\Delta ^{++}-2\Delta ^{+}+\Delta ^0=0. \label{sr10}
\end{equation}

Moreover, we can calculate $t$ from the following differences:
\begin{equation} \label{t}
t=\frac 13\left( \Omega ^{-}-\Delta ^{++}\right) =\frac 12\left( \Xi
^{*-}-\Delta ^{+}\right) =\Xi ^{*0}-\Sigma ^{*+}=\Sigma ^{*-}-\Delta ^0.
\end{equation}
These relations are interesting since they are related to the octet
splittings:%
$$
\Delta ^{++}-\Omega ^{-}=\frac 34\left( 2\text{p}+\text{n}+\Sigma
^{+}-\Sigma ^{-}-\Xi ^0-2\Xi ^{-}\right) .
$$
Experimentally: 6.46$\pm $0.59=7.38, where we have neglected errors on the
right hand side. As already remarked in the Introduction the error on the
left hand side which corresponds to the recent measurement of
Ref.\cite{Bosshard} would
dramatically increase if the world average for $\mu _{\Delta ^{++}}$ would
have been used.

Finally, let us present the formulae for the decuplet magnetic moments as
functions of $p$.  Because of very weak dependence on $p$ of the only
measured magnetic moments, namely $\Delta ^{++}$ and $\Omega ^{-}$, $p$
cannot be constrained from the existing decuplet data. Therefore in the
second and the third column of Table II we present two limiting cases
corresponding to the limits on $p$ somewhat extended with respect to
Eq.(\ref{plimits})\footnote{For example, the mixing parameter
$c_{\bar{10}}$ used in Ref.~\cite{Anti10} corresponds to $p=0.232$.}.

\subsection{Model predictions for the decuplet}

In Table III, the numerical results of the decuplet magnetic moments,
based on a selfconsistent solitonic profile function, are
presented. We can compare now our final results with the experimental data
for $\mu _{\Delta ^{++}}$ and $\mu _\Omega $. It is found that the present
model is in a remarkable agreement with the data within about $10\,\%$.
 The $m_{\rm s}$ corrections are by no means negligible.  Their largest
contribution is about $30\%$ and occurs in the $\mu _{\Delta ^{-}}$.

Similar conclusions have been drawn in Ref.\cite{WaKaya}, where both
octet and decuplet magnetic moments have been calculated in the model
identical to ours. There are however small quantitative differences
about which we would like to comment. The authors of Ref.\cite{WaKaya}
present their results for the constituent quark mass of 400~MeV with
a two-step regularization function as in Ref.\cite{Blotzetal}, however
with slightly different parameters. They also use soliton profile
function with massive pion tail corresponding to the non-strange quark
mass of 6~MeV. With these parameters in the chiral limit  they get
results which are only slightly different from ours, namely they get
for $t_{\rm chiral}=-2.28$ (see Eq.(\ref{decup})), whereas we get
$-2.34$. For $m_{\rm s}\ne 0$ the differences are, however, larger. In
our case magnetic moments of $\Delta^{++}$ and $\Omega^{-}$ increase
with $m_{\rm s}$ in absolute value, whereas in Ref.\cite{WaKaya} they
decrease. Moreover the magnetic moment of $\Xi^{*-}$ is independent of
$m_{\rm s}$ in \cite{WaKaya}, whereas in our case it increases in
absolute value.  We have explicitly checked that the results of 
Ref.~\cite{WaKaya} fulfill the sum rules of Eqs.(\ref{sr10}) and (\ref{t}).

Similar discrepancies have been found by authors of Ref.\cite{WaKaya}
in connection with the previous calculations of the octet magnetic
moments \cite{KimPolBloGo} by two of the present authors. They
attributed these discrepancies to the differences in the soliton
profile; this is, however, not correct since in both cases the massive
pion tail has been used. The differences in the decuplet case are
merely a reflection of the same problem to which at the moment  we do
not see any straightforward explanation.

\section{Summary and Conclusions}

In the present paper the magnetic moments of baryons have been studied
within the chiral quark soliton model with special emphasis on the
decuplet of baryons. We have adopted the strategy in which the model
has served as a tool to identify symmetry breaking terms proportional to
$m_{\rm s}$. Two sources of such terms are present in the model:
({\em i}) $m_{\rm s}$ corrections to the magnetic moment operator and
({\em ii}) wave function corrections which involve higher SU(3) flavor
representations. Having worked in perturbation theory in the symmetry
breaker we have derived analytical formulae for magnetic moments both
for the octet and decuplet.

For completeness also rotational $1/N_{\rm c}$ corrections have been
included.
They are important for model calculations with the explicit solitonic
profile function, however, they do not introduce any new algebraic
structure in the general expressions for the magnetic moments.

Symmetry breaking pattern for the magnetic moments is much more
involved than the one for the baryon masses, where Gell-Mann--Okubo
mass formulae work very well with only 2 free parameters. Here the
number of free parameters in the chiral limit is already 3 and one has
3 more if one goes off the chiral limit. Two more constants enter if one
takes into account wave function corrections. It is possible to
constrain all but one, called $p$, free parameters from the mass spectrum
and octet magnetic moments. It is interesting that $p$
is related to the splitting between well established octet or
decuplet and exotic baryon multiplets such as spin 1/2 antidecuplet
~\cite{Anti10}.
Unfortunately the only two known decuplet magnetic moments, namely that
of $\Delta^{++}$ and $\Omega^-$, do not provide any handle on this
splitting, since they are almost independent of $p$. In this respect precise
measurements of $\Delta^0$, $\Delta^-$ or $\Sigma^{*-}$ could
serve as a tool to extract $p$ and, in consequence, the masses of the
exotic states.

We have also found sum rules which the magnetic moments satisfy
in this order of the perturbation theory.

Finally we have presented model results for  the decuplet magnetic
moments. Our results agree with the existing data reasonably well,
however, as far as the value of $w_3$ is concerned,
there is for the value of $w_3$ substantial 
difference between the numerics of our ``model-independent'' analysis 
(which merely uses the algebraic structure of the $\chi$QSM) 
and actual model calculations, based on selfconsistent profile functions.
 This is reflected in the sign difference between
``model-independent'' predictions and model results for $\Delta^0$ and
$\Sigma^{*0}$. In general the two sets of predictions agree
each other within $20\%$. At this point it is important
to stress that this relatively good
agreement could be achieved only by taking into account rotational
$1/N_{\rm c}$ corrections. This can be clearly seen from Table III.
Similarly to the axial constants~\cite{BloPra},
magnetic moments would be factor of 2
off experimental data if these corrections were discarded.

We have compared our results with those of Ref.\cite{WaKaya}. In the
chiral limit the agreement between  two papers is quite satisfactory.
However, we have found some numerical differences for the finite strange
quark mass. Since the parameters and numerical procedures used in both
cases are very similar these differences are somewhat surprising and
require further study.

Let us stress at the end that parameters $w_i$ extracted from the data
in the ``model-independent'' analysis may be used to predict some of the
constants governing semileptonic hyperon decays~\cite{KimPoPraGo}.

\section*{Acknowledgment}

We would like to thank M.V. Polyakov for fruitful discussions and critical
comments.  We are grateful to T.A. Kaeding for providing us with the code
of SU(3) coupling coefficients.
This work has partly been supported by the BMBF, the DFG and the
COSY--Project (J\" ulich).  M.P. acknowledges support of Alexander
von Humboldt Stiftung.

\section*{Appendix}
In this appendix, we present all formulae appearing in Eq.(\ref{Eq:mu}).
\begin{eqnarray}
I_1 & = & \frac{N_c}{6} \sum_{n}
\int d^3 x \;\int d^3 y
\left [\frac{\Psi^{\dagger}_{n} ({\bf x}) {\tau}
\Psi_{val} ({\bf x}) \cdot
\Psi^{\dagger}_{val} ({\bf y}) {\tau} \Psi_{n} ({\bf y})}
{E_n - E_{val}} \right .
\nonumber \\  & & \hspace{3cm} \;+\; \left .
\frac{1}{2}\sum_{m}
\Psi^{\dagger}_{n} ({\bf x}) {\tau} \Psi_{m} ({\bf x}) \cdot
\Psi^{\dagger}_{m} ({\bf y}) {\tau} \Psi_{n} ({\bf y})
{\cal R}_{\cal I} (E_n, E_m) \right ],
\nonumber \\
I_2 & = &\frac{N_c}{6} \sum_{m^{0}}
\int d^3 x \;\int d^3 y
\left [\frac{\Psi^{\dagger}_{m^{0}} ({\bf x}) \Psi_{val} ({\bf x})
\Psi^{\dagger}_{val} ({\bf y}) \Psi_{m{^0}} ({\bf y})}
{E_{m^{0}} - E_{val}} \right .
\nonumber \\  & & \hspace{3cm} \;+\;\left . \frac{1}{2}\sum_{n}
\Psi^{\dagger}_{n} ({\bf x}) \Psi_{m^{0}} ({\bf x})
\Psi^{\dagger}_{m^{0}} ({\bf y}) \Psi_{n} ({\bf y})
{\cal R}_{\cal I} (E_n, E_m^{0}) \right ],
\nonumber \\
K_1 & = & \frac{N_c}{6} \sum_{n}
\int d^3 x \; \int d^3 y
\left [\frac{\Psi^{\dagger}_{n} ({\bf x})
{\tau} \Psi_{val} ({\bf x}) \cdot
\Psi^{\dagger}_{val} ({\bf y}) \beta {\tau} \Psi_{n} ({\bf y})}
{E_n - E_{val}} \right .
\nonumber \\  & & \hspace{3cm} \;+\; \left . \frac{1}{2}\sum_{m}
\Psi^{\dagger}_{n} ({\bf x}) {\tau} \Psi_{m} ({\bf x}) \cdot
\Psi^{\dagger}_{m} ({\bf y}) \beta {\tau} \Psi_{n} ({\bf y})
{\cal R}_{\cal M} (E_n, E_m)
\right ],
\nonumber \\
K_2 & = & \frac{N_c}{6} \sum_{m^{0}}
\int d^3 x \; \int d^3 y
\left [\frac{\Psi^{\dagger}_{m^{0}} (x) \Psi_{val} ({\bf x})
\Psi^{\dagger}_{val} ({\bf y}) \beta \Psi_{m{^0}} ({\bf y})}
{E_{m^{0}} - E_{val}} \right .
\nonumber \\  & & \hspace{3cm} \;+\;\left . \frac{1}{2}\sum_{n}
\Psi^{\dagger}_{n} ({\bf x}) \Psi_{m^{0}} ({\bf x})
\Psi^{\dagger}_{m^{0}} ({\bf y}) \beta \Psi_{n} ({\bf y})
{\cal R}_{\cal M} (E_n, E_m^{0}) \right ],\\
{\cal Q}_0  & = &
\frac{N_c}{3}\int d^3 x \,
\left[ \Psi ^{\dagger}_{val}({\bf x}) \gamma_{5}
\{{\bf r} \times {\sigma} \} \cdot {\tau}
\Psi_{val} ({\bf x}) \right. \nonumber \\ & &
\left . \hspace{1cm} \;-\;
\frac{1}{2}  \sum_n {\rm sgn} (E_n)
\Psi ^{\dagger}_{n}({\bf x}) \gamma_{5}
\{{\bf r} \times {\sigma} \} \cdot {\tau}
\Psi_{n}({\bf x}) {\cal R}(E_n)\right ],
\nonumber \\
{\cal Q}_1 & = &  \frac{iN_c}{6}\sum_{n}
\int d^3 x \,
\int d^3 y \nonumber \\  & \times &
\left[{\rm sgn} (E_n)
\frac{\Psi^{\dagger}_{n} ({\bf x}) \gamma_{5}
\{{\bf r} \times {\sigma} \} \times {\tau}
\Psi_{val} ({\bf x}) \cdot
\Psi^{\dagger}_{val} ({\bf y}) {\tau} \Psi_{n} ({\bf y})}
{E_n - E_{val}} \right .
\nonumber \\  & & 
\;+\; \left . \frac{1}{2} \sum_{m}
\Psi^{\dagger}_{n} ({\bf x})\gamma_{5} \{{\bf r} \times {\sigma} \}
\times {\tau}  \Psi_{m} ({\bf x}) \cdot
\Psi^{\dagger}_{m} ({\bf y}) {\tau} \Psi_{n} ({\bf y})
{\cal R}_{\cal Q} (E_n, E_m) \right ],
\nonumber \\
{\cal Q}_2  & = &  \frac{N_c}{6} \sum_{m^0}
\int d^3 x \,
\int d^3 y \nonumber \\  & \times &
\left[ {\rm sgn} (E_{m^0})
\frac{\Psi^{\dagger}_{m^0} ({\bf x}) \gamma_{5}
\{{\bf r} \times {\sigma} \} \cdot {\tau}
\Psi_{val} ({\bf x})
\Psi^{\dagger}_{val} ({\bf y}) \Psi_{m^0} ({\bf y})}
{E_{m^0} - E_{val}} \right .
\nonumber \\  & + & 
\left . \sum_{n}
\Psi^{\dagger}_{n} ({\bf x})\gamma_{5} \{{\bf r} \times {\sigma} \}
\cdot {\tau}  \Psi_{m^0} ({\bf x})
\Psi^{\dagger}_{m^0} ({\bf y}) \Psi_{n} ({\bf y})
{\cal R}_{\cal Q} (E_n, E_{m^0}) \right ],
\nonumber \\
{\cal X}_1  & = & \frac{N_c}{3}
\sum_{n} \int d^3 x \,
\int d^3 y \left[
\frac{\Psi^{\dagger}_{n} ({\bf x})\gamma_{5}
\{{\bf r} \times {\sigma} \}
\Psi_{val} ({\bf x}) \cdot
\Psi^{\dagger}_{val} ({\bf y}) {\tau} \Psi_{n} ({\bf y})}
{E_n - E_{val}} \right .
\nonumber \\  &+ & \left. 
\frac{1}{2} \sum_{m}
\Psi^{\dagger}_{n} ({\bf x})\gamma_{5} \{{\bf r} \times {\sigma} \}
\Psi_{m} ({\bf x}) \cdot
\Psi^{\dagger}_{m} ({\bf y}) {\tau} \Psi_{n} ({\bf y})
{\cal R}_{\cal M} (E_n, E_m) \right ],
\nonumber \\
{\cal X}_2  & = & \frac{N_c}{3}
\sum_{m^0} \int d^3 x \,
\int d^3 y \left[
\frac{\Psi^{\dagger}_{m^0} ({\bf x})\gamma_{5}
\{{\bf r} \times {\sigma} \} \cdot {\tau}
\Psi_{val} ({\bf x})
\Psi^{\dagger}_{val} ({\bf y}) \Psi_{m^0} ({\bf y})}
{E_{m^0} - E_{val}} \right .
\nonumber \\  & & \hspace{1cm}
\;+\; \left . \sum_{n}
\Psi^{\dagger}_{n} ({\bf x})\gamma_{5} \{{\bf r} \times {\sigma} \}
\cdot {\tau} \Psi_{m^0} ({\bf x})
\Psi^{\dagger}_{m^0} ({\bf y}) \Psi_{n} ({\bf y})
{\cal R}_{\cal M} (E_n, E_{m^0}) \right ],
\nonumber \\
{\cal M}_0  & = &  \frac{N_c}{9} \sum_{n}
\int d^3 x \,
\int d^3 y \left[
\frac{\Psi^{\dagger}_{n} ({\bf x}) \gamma_{5}
\{{\bf r} \times {\sigma} \} \cdot {\tau}
\Psi_{val} ({\bf x})
\Psi^{\dagger}_{val} ({\bf y}) \beta \Psi_{n} ({\bf y})}
{E_{n} - E_{val}} \right .
\nonumber \\  & & \hspace{1cm}
\;+\; \left . \frac{1}{2} \sum_{m}
\Psi^{\dagger}_{n} ({\bf x})\gamma_{5} \{{\bf r} \times {\sigma} \}
\cdot {\tau}  \Psi_{m} ({\bf x})
\Psi^{\dagger}_{m} ({\bf y})\beta \Psi_{n} ({\bf y})
{\cal R}_{\beta} (E_n, E_m) \right ],
\nonumber \\
{\cal M}_1  & = & \frac{N_c}{9}
\sum_{n} \int d^3 x \,
\int d^3 y \nonumber \\  & \times &
\left[
\frac{\Psi^{\dagger}_{n} ({\bf x})\gamma_{5}
\{{\bf r} \times {\sigma} \}
\Psi_{val} ({\bf x}) \cdot
\Psi^{\dagger}_{val} ({\bf y}) \beta {\tau} \Psi_{n} ({\bf y})}
{E_n - E_{val}} \right .
\nonumber \\  &+ & 
\left . \frac{1}{2} \sum_{m}
\Psi^{\dagger}_{n} ({\bf x})\gamma_{5} \{{\bf r} \times {\sigma} \}
\Psi_{m} ({\bf x}) \cdot
\Psi^{\dagger}_{m} ({\bf y}) \beta {\tau} \Psi_{n} ({\bf y})
{\cal R}_{\beta} (E_n, E_m) \right ],
\nonumber \\
{\cal M}_2 & = &  \frac{N_c}{9} \sum_{m^0}
\int d^3 x \,
\int d^3 y \nonumber \\  & \times &
\left[
\frac{\Psi^{\dagger}_{m^0} ({\bf x}) \gamma_{5}
\{{\bf r} \times {\sigma} \} \cdot {\tau}
\Psi_{val} ({\bf x})
\Psi^{\dagger}_{val} ({\bf y})\beta \Psi_{m^0} ({\bf y})}
{E_{m^0} - E_{val}} \right .
\nonumber \\  &+ & 
\left . \sum_{n}
\Psi^{\dagger}_{n} ({\bf x})\gamma_{5} \{{\bf r} \times {\sigma} \}
\cdot {\tau}  \Psi_{m^0} ({\bf x})
\Psi^{\dagger}_{m^0} ({\bf y}) \beta \Psi_{n} ({\bf y})
{\cal R}_{\beta} (E_n, E_{m^0}) \right ]  .
\label{Eq:mdens}
\end{eqnarray}
The regularization functions in Eq.(\ref{Eq:mdens}) are as follows:
\begin{eqnarray}
{\cal R}_{I} (E_n, E_m) & = & - \frac{1}{2\sqrt{\pi}}
\int^{\infty}_{0} \frac{du}{\sqrt{u}} \phi (u;\Lambda_i)
\left [ \frac{E_n e^{-u E^{2}_{n}} +  E_m e^{-u E^{2}_{m}}}
{E_n + E_m} \;+\; \frac{e^{-u E^{2}_{n}} - e^{-u E^{2}_{m}}}
{u(E^{2}_{n} - E^{2}_{m})} \right ],
\nonumber \\
{\cal R}_{\cal M} (E_n, E_m) & = &
\frac{1}{2}  \frac{ {\rm sgn} (E_n)
- {\rm sgn} (E_m)}{E_n - E_m},\nonumber \\
{\cal R} (E_n) & = & \int \frac{du}{\sqrt{\pi u}}
\phi (u;\Lambda_i) |E_n| e^{-uE^{2}_{n}},
\nonumber \\
{\cal R}_{\cal Q} (E_n, E_m) & = & \frac{1}{2\pi} c_i
\int^{1}_{0} d\alpha \frac{\alpha (E_n + E_m) - E_m}
{\sqrt{\alpha ( 1 - \alpha)}}
\frac{\exp{\left (-[\alpha E^{2}_n + (1-\alpha)E^{2}_m]/
\Lambda^{2}_i  \right)}}{\alpha E^{2}_n + (1-\alpha)E^{2}_m},
\nonumber \\
{\cal R}_{\beta}  (E_n, E_m) & = &
\frac{1}{2\sqrt{\pi}} \int^{\infty}_{0}
\frac{du}{\sqrt{u}} \phi (u;\Lambda_i)
\left[ \frac{E_n e^{-uE^{2}_{n}} - E_m e^{-uE^{2}_{m}}}
{E_n - E_m}\right],
\label{Eq:regulm}
\end{eqnarray}
where the cutoff function $\phi(u;\Lambda_i)=\sum_i c_i \theta
\left(u - \frac{1}{\Lambda^{2}_{i}} \right)$ is
fixed by reproducing the pion decay constant and other mesonic properties
\cite{review}.

\vfill\eject
\begin{table}[]
\caption{The dependence of the magnetic moments of the SU(3) decuplet
baryons on the constituent quark mass $M$ without $m_{{\rm s}}$ corrections:
$\mu (\Omega^0)$ corresponds to the leading order in the rotational
frequency while $\mu (\Omega^1)$ includes the subleading order. The unit of
magnetic moments are given in nuclear magneton $\mu_N$.
 The numbers are obtained by using a selfconsistently calculated 
solitonic profile function.}
\begin{tabular}{|c||c|c|c|c|c|c|c|c|c|}
\multicolumn{1}{|c||}{\phantom{0}} & \multicolumn{2}{c|}{370 MeV} &
\multicolumn{2}{c|}{400 MeV} & \multicolumn{2}{c|}{420 MeV} &
\multicolumn{2}{c|}{450 MeV} & \multicolumn{1}{c|}{\phantom{0}} \\
\cline{2-9}
\multicolumn{1}{|c||}{Baryon} & \multicolumn{1}{c|}{$\mu_B (\Omega^0)$} &
\multicolumn{1}{c|}{$\mu_B (\Omega^1)$} & \multicolumn{1}{c|}{$\mu_B
(\Omega^0)$} & \multicolumn{1}{c|}{$\mu_B (\Omega^1)$} & \multicolumn{1}{c|}{$\mu_B (\Omega^0)$} & \multicolumn{1}{c|}{$\mu_B (\Omega^1)$} &
\multicolumn{1}{c|}{$\mu_B (\Omega^0)$} & \multicolumn{1}{c|}{$\mu_B
(\Omega^1)$} & \multicolumn{1}{c|}{Exp} \\ \hline
$\Delta^{++}$ & $2.11$ & $5.07$ & $1.96$ & $4.67$ & $1.88$ & $4.47$ & $1.68$
& $4.16$ & $4.52\pm 0.5$ \\
$\Delta^{+}$ & $1.05$ & $2.53$ & $0.98$ & $2.34$ & $0.94$ & $2.23$ & $0.84$
& $2.08$ & $--$ \\
$\Delta^{0}$ & $0$ & $0$ & $0$ & $0$ & $0$ & $0$ & $0$ & $0$ & $--$ \\
$\Delta^{-}$ & $-1.05$ & $-2.53$ & $-0.98$ & $-2.34$ & $-0.94$ & $-2.23$ & $%
-0.84$ & $-2.08$ & $--$ \\
$\Sigma^{*+}$ & $1.05$ & $2.53$ & $0.98$ & $2.34$ & $0.94$ & $2.23$ & $0.84$
& $2.08$ & $--$ \\
$\Sigma^{*0}$ & $0$ & $0$ & $0$ & $0$ & $0$ & $0$ & $0$ & $0$ & $--$ \\
$\Sigma^{*-}$ & $-1.05$ & $-2.53$ & $-0.98$ & $-2.34$ & $-0.94$ & $-2.23$ & $%
-0.84$ & $-2.08$ & $--$ \\
$\Xi^{*0}$ & $0$ & $0$ & $0$ & $0$ & $0$ & $0$ & $0$ & $0$ & $--$ \\
$\Xi^{*-}$ & $-1.05$ & $-2.53$ & $-0.98$ & $-2.34$ & $-0.94$ & $-2.23$ & $%
-0.84$ & $-2.08$ & $--$ \\
$\Omega^{-}$ & $-1.05$ & $-2.53$ & $-0.98$ & $-2.34$ & $-0.94$ & $-2.23$ & $%
-0.84$ & $-2.08$ & $-1.94 \pm 0.31$ \\
\end{tabular}
\end{table}
\begin{table}[]
\caption{``Model-independent'' results to linear order in $m_{\rm s}$
for decuplet magnetic moments as
functions of $p$ defined in Eq.(\protect\ref{Eq:pq}).
 The unit of magnetic moments are given in nuclear magneton $\mu_N$.}
\begin{tabular}{ccccc}
Baryons & & $p=0.15$ & $p=0.25$ & Exp \\ \hline
$\Delta^{++}$ & $5.322 + 0.087  \, p$ & $5.33 $ & $5.34$  & $4.52\pm 0.5$\\ 
$\Delta^{+}$  & $2.847 - 0.715  \, p$ & $2.74 $ & $2.67$  & $--$ \\
$\Delta^{0}$  & $0.371 - 1.516  \, p$ & $0.14 $ & $-0.01$ & $--$ \\
$\Delta^{-}$  & $-2.104- 2.318  \, p$ & $-2.45$ & $-2.68$ & $--$ \\ 
$\Sigma^{*+}$ & $2.987 + 0.442  \, p$ & $3.05 $ & $ 3.10$ & $--$ \\
$\Sigma^{*0}$ & $0.449 - 0.537  \, p$ & $0.37 $ & $ 0.32$ & $--$ \\
$\Sigma^{*-}$ & $-2.089 - 1.516 \, p$ & $-2.32$ & $-2.47$ & $--$ \\
$\Xi^{*0}$    & $0.527 + 0.442  \, p$ & $0.59 $ & $ 0.64$ & $--$ \\
$\Xi^{*-}$    & $-2.054 - 0.715 \, p$ & $-2.18$ & $-2.25$ & $--$ \\
$\Omega^{-}$  & $-2.060 + 0.087 \, p$ & $-2.05$ & $-2.04$ & $-1.94 \pm 0.31$\\
\end{tabular}
\end{table}
\begin{table}[]
\caption{The magnetic moments of the SU(3) decuplet baryons with $m_{{\rm s}}
$ corrections: $\mu (\Omega^0,m^0_{\rm s})$ corresponds to the leading 
order in the rotational frequency while $\mu (\Omega^1,m^0_{\rm s})$ 
includes the subleading order. $\mu (\Omega^1, m^1_{s})$ represents 
our final results.
 The constituent quark mass $M$ is fixed to be $420$ MeV.  The unit of
magnetic moments are given in nuclear magneton $\mu_N$.
The numbers are obtained by using a selfconsistently calculated 
solitonic profile function.}
\begin{tabular}{ccccc}
Baryons & $\mu_B (\Omega^0, m^{0}_{\rm s})$ & 
$\mu_B(\Omega^1, m^{0}_{\rm s})$ & $%
\mu_B(\Omega^1, m^{1}_{\rm s})$ & $\mbox{Exp.}$ \\ \hline
$\Delta^{++}$ & $1.88 $ & $4.47 $ & $4.73 $ & $4.52\pm0.50$ \\
$\Delta^{+}$ & $0.94 $ & $2.23 $ & $2.19 $ & --- \\
$\Delta^{0}$ & $0 $ & $0 $ & $-0.35$ & --- \\
$\Delta^{-}$ & $-0.94$ & $-2.23$ & $-2.90$ & --- \\
$\Sigma^{*+}$ & $0.94 $ & $2.23 $ & $2.52 $ & --- \\
$\Sigma^{*0}$ & $0 $ & $0 $ & $-0.08$ & --- \\
$\Sigma^{*-}$ & $-0.94$ & $-2.23$ & $-2.69$ & --- \\
$\Xi^{*0}$ & $0 $ & $0 $ & $0.19 $ & --- \\
$\Xi^{*-}$ & $-0.94$ & $-2.23$ & $-2.48$ & --- \\
$\Omega^{-}$ & $-0.94$ & $-2.23$ & $-2.27$ & $-1.94\pm 0.31$ \\
\end{tabular}
\end{table}

\end{document}